\documentclass[aps,pre,floatfix,twocolumn,groupedaddress]{revtex4-1}
\usepackage[dvips]{graphicx}
\usepackage{amsmath}
\usepackage{verbatim}
\newcommand{\etalia}{{\it et al.~}}
\newcommand{\la}{\left\langle}
\newcommand{\ra}{\right\rangle}

\newcommand{\PRL}{Phys.~Rev.~Lett.~}
\newcommand{\PR}{Phys.~Rev.~}
\newcommand{\JCP}{J.~Chem.~Phys.~}

\newcommand{\JPC}{J.~Phys.~Chem.~}
\newcommand{\JPCM}{J.~Phys.: Condens.~Matter~}

\begin{document}

\title{Polymer Crowding and Shape Distributions in Polymer-Nanoparticle Mixtures}

\author{Wei Kang Lim and Alan R. Denton}
\email[]{alan.denton@ndsu.edu}
\affiliation{Department of Physics, North Dakota State University,
Fargo, ND 58108-6050, USA}

\begin{abstract}
Macromolecular crowding can influence polymer shapes, which is important for 
understanding the thermodynamic stability of polymer solutions and the
structure and function of biopolymers (proteins, RNA, DNA) under confinement.
We explore the influence of nanoparticle crowding on polymer shapes via
Monte Carlo simulations and free-volume theory of a coarse-grained model
of polymer-nanoparticle mixtures.  Exploiting the geometry of random walks,
we model polymer coils as effective penetrable ellipsoids, whose shapes
fluctuate according to the probability distributions of the eigenvalues of 
the gyration tensor.  Accounting for the entropic cost of a nanoparticle
penetrating a larger polymer coil, we compute the crowding-induced shift in
the shape distributions, radius of gyration, and asphericity of ideal polymers
in a theta solvent.
With increased nanoparticle crowding, we find that polymers become more compact
(smaller, more spherical), in agreement with predictions of free-volume theory.
Our approach can be easily extended to nonideal polymers in good solvents and 
used to model conformations of biopolymers in crowded environments.
\end{abstract}

\maketitle
\newpage


\section{Introduction}
Polymers are commonly confined within biological systems and other soft materials~\cite{minton2001}.
Confinement can result from geometric boundaries, as in thin films and porous media,
or from crowding by other species, as in nanocomposite materials and cellular environments.
Within the nucleoplasm and cytoplasm of eukaryotic cells, for example,
an assortment of macromolecules (proteins, RNA, DNA, etc.) share a tightly restricted space, 
occupying from 20\% to 40\% of the total volume~\cite{vandermaarel2008,phillips2009}.
In this crowded milieu, smaller molecules exclude volume to larger, softer biopolymers,
constraining conformations and influencing folding pathways.  Macromolecular crowding,
because of its profound influence on the structure, and hence function, of biopolymers,
has been intensely studied over the past three decades~\cite{minton1981,minton2000,minton2005,
ellis2001,richter2007,richter2008,elcock2010,hancock2012,denton-cmb2013}.

It is well established that crowding can significantly modify polymer conformations.
The asymmetric shapes of folded and denatured states of biopolymers, in particular,
are known to respond sensitively to the presence of crowding agents~\cite{goldenberg2003,
dima2004,cheung2005,wittung-stafshede2012,linhananta2012,denesyuk2011,denesyuk2013}.  
The shape distribution of a protein or RNA, for example, can vary with crowder
concentration, which in turn, can affect the biopolymer's function.
Polymer shapes are also important in determining the nature of depletion-induced
effective interactions between colloids and nanoparticles, thereby influencing
thermodynamic stability of colloid-polymer mixtures against demixing.
Direct measurements~\cite{yodh2001} show, for example, that rodlike and spherical
depletants induce significantly different interactions between colloids.
Confinement and crowding effects are thus of practical concern for 
their impact on the properties of polymer-nanoparticle composite materials~\cite{han2001,
kramer2005a,kramer2005b,kramer2005c,balazs2006,mackay2006,richter2010,denton-cmb2013}
and for their role in diseases associated with protein aggregation~\cite{stradner2007}. 

Fundamental interest in polymer shapes dates to the dawn of polymer science.
Already 80 years ago, Kuhn~\cite{kuhn1934} recognized that macromolecules in solution
are fluctuating objects, whose shapes are far from spherical, and that a linear polymer
chain, when viewed in its principal-axis frame of reference, resembles a significantly
elongated, flattened (bean-shaped) ellipsoid.  The close analogy between polymers and
random walks has inspired many mathematical and statistical mechanical studies to analyze
sizes and shapes of random walks~\cite{fixman1962,flory-fisk1966,flory1969,yamakawa1970,
fujita1970,solc1971,solc1973,theodorou1985,rudnick-gaspari1986,rudnick-gaspari1987,
bishop1988,sciutto1996,murat-kremer1998,eurich-maass2001}.
Such studies validate Kuhn's insight and reveal broad distributions of radius of gyration
and shape, as characterized by the eigenvalues of the gyration tensor.

In the case of colloidal particles larger than polymer radii of gyration
(colloid limit), polymer depletion and induced effective attraction
between colloids are relatively well understood
phenomena~\cite{asakura1954,vrij1976,pusey1991,jones2002,fuchs2002}.
The opposite case, in which smaller colloids (nanoparticles) can easily
penetrate larger polymers (protein limit), has been studied more recently
by theory~\cite{sear1997,sear2001,sear2002}, 
simulation~\cite{bolhuis2002,bolhuis2003,moncho-jorda2003,cheung2013}, and
experiment~\cite{vanduijneveldt2005,vanduijneveldt2006,vanduijneveldt2007}.
Previous studies, while analyzing depletion-induced interactions and 
demixing phase behavior, have not directly addressed the response of 
polymer shape to crowding.
The purpose of this paper is to explore the influence of nanoparticle crowding
on the shapes of polymers in polymer-nanoparticle mixtures.

In the next section, we define our model of a polymer-nanoparticle mixture.
In Sec.~\ref{methods}, we describe our simulation method and outline the 
free-volume theory, relegating details to an appendix.
In Sec.~\ref{results}, we present results from our simulations for the
shape distributions of crowded polymers and compare with theoretical predictions.
Finally, in Sec.~\ref{conclusions}, we summarize and suggest possible extensions
of our approach for future work.

\section{Models}\label{models}
\subsection{Polymer-Nanoparticle Mixtures}\label{mixtures}
We model a mixture of nanoparticles and nonadsorbing polymers using a 
generalization of the Asakura-Oosawa-Vrij (AOV) model of 
colloid-polymer mixtures~\cite{asakura1954,vrij1976}.
The original AOV model represents the particles as hard (impenetrable) spheres,
interacting via a hard-sphere pair potential,
\begin{eqnarray}
v_{nn}(r)&=&\left\{\begin{array}{l@{\quad\quad}l}
\infty~, \qquad & r<2R_n~, \\[1ex]
0~, \qquad & r\geq 2R_n~,
\end{array} \right.
\end{eqnarray}
and the polymers as effective spheres of fixed size (radius of gyration) 
that are mutually ideal (noninteracting), but impenetrable to the particles.
While the neglect of polymer-polymer interactions is justified for polymers 
in a theta solvent~\cite{degennes1979}, the effective-sphere approximation
ignores aspherical conformations and shape fluctuations of polymer coils.
Moreover, the assumption of hard polymer-particle interactions is physically
reasonable only for particles much larger than the polymers.
In order to study the influence of nanoparticle crowding on polymer shapes, 
we generalize the AOV model by allowing nanoparticles to penetrate polymers
and by representing the polymers as ellipsoids that fluctuate in size and shape.
\begin{figure}
\begin{center}
\includegraphics[width=0.8\columnwidth]{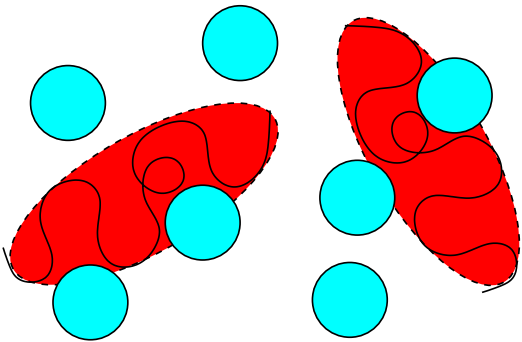}
\end{center}
\caption{Model of polymer-nanoparticle mixtures.  Polymers are penetrable ellipsoids
that can fluctuate in size and shape.  Nanoparticles are hard spheres of fixed size.
}\label{fig-model}
\end{figure}

Following Schmidt and Fuchs~\cite{schmidt-fuchs2002}, we attribute to each 
overlapping polymer-nanoparticle pair, an average free energy cost $\varepsilon$, 
which accounts for the loss in conformational entropy of the coil.  
For a hard sphere penetrating an ideal polymer coil,
in a theta solvent at temperature $T$, polymer field theory predicts
$\varepsilon=3k_BT/q$, where $q=R_p/R_n$ is the ratio of the polymer radius of
gyration $R_p$ to the nanoparticle radius $R_n$~\cite{eisenriegler1996,hanke1999}.
An obvious refinement of this model would allow the overlap free energy
to vary with the nanoparticle's position relative to the polymer center.
Such effective interaction energy profiles have been computed from Monte Carlo
simulations of polymers on a lattice~\cite{pelissetto-hansen2006}.
Alternatively, the overlap free energy profile could be derived from 
an approximation for the monomer density in the ellipsoidal polymer 
model~\cite{eurich-maass2001} (see below).
In the current study, however, for conceptual simplicity and computational
efficiency, we neglect this level of spatial resolution.
Furthermore, since the nanoparticles in our model are chemically inert and
act only to limit the free volume available to the polymers, we assume that
the theta temperature of the solution is independent of nanoparticle concentration.

\subsection{Penetrable Polymer Model}\label{penetrable-polymer}
The size and shape of a polymer coil can be characterized by the gyration tensor,
defined by
\begin{equation}
{\bf T}~=~\frac{1}{N}\sum_{i=1}^N{\bf r}_i~{\bf r}_i~,
\label{gyration-tensor}
\end{equation}
where ${\bf r}_i$ denotes the position of the $i^{\rm th}$ of $N$ segments,
relative to the center of mass.
Any particular conformation has a radius of gyration defined by
\begin{equation}
R_p=\left(\frac{1}{N}\sum_{i=1}^Nr_i^2\right)^{1/2}=\sqrt{\Lambda_1+\Lambda_2+\Lambda_3}~,
\label{radius-gyration}
\end{equation}
where $\Lambda_1$, $\Lambda_2$, $\Lambda_3$ are the eigenvalues of ${\bf T}$.
For reference, the gyration tensor is related to the moment of inertia tensor
${\bf I}$, familiar from classical mechanics of rigid bodies, via 
${\bf T}=R_p^2{\bf 1}-{\bf I}$, where ${\bf 1}$ is the unit tensor.
The root-mean-square (rms) radius of gyration, which is experimentally measurable,
is given by
\begin{equation}
R_g=\sqrt{\la R_p^2\ra}=\sqrt{\la\Lambda_1+\Lambda_2+\Lambda_3\ra}~,
\label{gyration-avg}
\end{equation}
where the angular brackets represent an ensemble average over conformations.

Now, if the average in Eq.~(\ref{gyration-avg}) is defined relative to a 
fixed (laboratory) frame of reference, then the average tensor is symmetric, 
has equal eigenvalues, and describes a sphere.  If instead the average is
performed in a frame of reference that rotates with the polymer's principal axes,
the coordinate axes being labelled to preserve the order of the eigenvalues by
magnitude ($\Lambda_1>\Lambda_2>\Lambda_3$), then the average tensor is asymmetric
and describes an anisotropic object~\cite{rudnick-gaspari1986,rudnick-gaspari1987}.
In other words, viewed from the laboratory frame, the average shape of a 
random walk is spherical, but viewed from the principal-axis frame, the 
average shape is aspherical~\cite{kuhn1934}.
In fact, in the principal-axis frame, the average shape is a significantly 
elongated (prolate), flattened ellipsoid with principal radii 
along the three independent axes in the approximate ratio 
3.4~:~1.6~:~1~\cite{kuhn1934,solc1971,solc1973}.
Each eigenvalue of the gyration tensor is proportional to the square of the
respective principal radius of the general ellipsoid that best fits the 
shape of the polymer,  an arbitrary point $(x,y,z)$ on the surface of the 
ellipsoid satisfying
\begin{equation}
\frac{x^2}{\Lambda_1}+\frac{y^2}{\Lambda_2}+\frac{z^2}{\Lambda_3}=3~.
\label{ellipsoid}
\end{equation}
This ellipsoid serves as a gross representation of the tertiary structure
of a biopolymer.

The shape of an ideal, freely-jointed polymer coil of $N$ segments of length $l$,
modeled as a soft Gaussian ellipsoid~\cite{murat-kremer1998}, has a normalized
probability distribution that is well approximated by the analytical ansatz of
Eurich and Maass~\cite{eurich-maass2001}:
\begin{equation}
P_r(\lambda_1,\lambda_2,\lambda_3) = \prod_{i=1}^3 P_{ir}(\lambda_i)~,
\label{Prlambda}
\end{equation}
where $\lambda_i\equiv\Lambda_i/(Nl^2)$ are scaled (dimensionless) eigenvalues and 
\begin{equation}
P_{ir}(\lambda_i) = \frac{(a_id_i)^{n_i-1}\lambda_i^{-n_i}}{2K_i}
\exp\left(-\frac{\lambda_i}{a_i}-d_i^2\frac{a_i}{\lambda_i}\right)~,
\label{Eurich-Maass}
\end{equation}
with fitting parameters 
$K_1=0.094551$, $K_2=0.0144146$, $K_3=0.0052767$, 
$a_1=0.08065$, $a_2=0.01813$, $a_3=0.006031$,
$d_1=1.096$, $d_2=1.998$, $d_3=2.684$, 
$n_1=1/2$, $n_2=5/2$, and $n_3=4$. 
The assumption of independent eigenvalues underlying the factorization ansatz
of Eq.~(\ref{Prlambda}) is not exact, since an extension of a 
random walk in one direction affects the probability of an extension in an 
orthogonal direction.  Nevertheless, conformations that significantly violate 
the ansatz are rare for random walks sufficiently long to model real polymers.
It should be noted that the ellipsoidal polymer model has also been extended
to block copolymers~\cite{eurich2007}.

In modeling mixtures of polymers and nanoparticles, it is convenient to consider
the system to be in osmotic equilibrium with a reservoir of pure polymer, 
which fixes the polymer chemical potential.  A key parameter that defines the system
is the ratio, $q_r\equiv R_g^r/R_n$, of the rms radius of gyration of polymer 
in the reservoir $R_g^r$ to the nanoparticle radius.
Expressed in terms of the scaled eigenvalues, the ratio of the rms radius of gyration
in the system [Eq.~(\ref{gyration-avg})] to its counterpart in the reservoir
[$R_g^r=l\sqrt{N/6}$] is given by
\begin{equation}
\frac{R_g}{R_g^r}=\sqrt{6\la\lambda_1+\lambda_2+\lambda_3\ra}~.
\label{gyration-ratio}
\end{equation}
Similarly, the principal radii are related to the scaled eigenvalues according to
\begin{equation}
R_i=R_g^r\sqrt{18\lambda_i}~, \quad i=1,2,3~.
\label{principal-radii}
\end{equation}

The broad eigenvalue distributions described by Eq.~(\ref{Eurich-Maass}) imply 
significant fluctuations in size ($R_g$) and shape ($\lambda_i$) of the polymer 
[see Fig.~(\ref{fig-Plambda-q5}) below].
The deviation of a polymer's average shape from a perfect sphere can be quantified
by an asphericity parameter~\cite{rudnick-gaspari1986,rudnick-gaspari1987}
\begin{equation}
A=1-3\frac{\la\lambda_1\lambda_2+\lambda_1\lambda_3+\lambda_2\lambda_3\ra}
{\la(\lambda_1+\lambda_2+\lambda_3)^2\ra}~.
\label{asphericity}
\end{equation}
By this definition, a spherical object with all eigenvalues equal has $A=0$,
while an elongated object, with one eigenvalue much larger than the other two,
has $A\simeq 1$.
In the next section, we describe computational methods for calculating the 
shape distribution, radius of gyration, and asphericity of polymers crowded
by nanoparticles.

\section{Computational Methods}\label{methods}
\subsection{Monte Carlo Simulations}\label{simulation}
To explore the influence of nanoparticle crowding on polymer conformations, we have
developed a Monte Carlo (MC) method for simulating mixtures of hard nanoparticles and
ideal polymers, whose uncrowded shape distribution follows Eq.~(\ref{Eurich-Maass}).
In the canonical ensemble, the temperature, particle numbers ($N_n$ nanoparticles,
$N_p$ polymers), and volume $V$ are fixed.  Trial moves include displacements of
nanoparticles and, for the polymers, displacements, rotations, {\it and} shape changes.
In the standard Metropolis algorithm~\cite{binder1995,frenkel-smit2001,binder-heermann2010},
a trial move from an old to a new configuration, due to displacement of any particle 
or rotation of a polymer, is accepted with probability
\begin{equation}
{\cal P}_{\rm config}({\rm old}\to{\rm new})
=\min\left\{\exp(-\beta\Delta U),~1\right\}~,
\label{disp-rot}
\end{equation}
where $\beta=1/(k_B T)$ and $\Delta U$ is the associated change in potential energy.

Overlaps of hard-sphere nanoparticles are easily detected and are, of course, 
automatically rejected.  Polymer-nanoparticle overlaps, on the other hand, 
are harder to identify, because of the nontrivial calculation required to determine 
the shortest distance between the surface of a sphere and that of a general 
ellipsoid~\cite{frenkel-mulder1985}.  To avoid the computational overhead of 
this calculation, we here restrict our investigations to cases in which the 
nanoparticles are much smaller than the rms radius of gyration of the polymers
($q_r\gg 1$).  In this limit, we can accurately approximate the volume excluded
by a polymer to a nanoparticle, whose true shape is an ellipsoid coated by a 
shell of uniform thickness $R_n$, by a larger ellipsoid, whose principal radii
are extended by $R_n$.  Thus, we approximate the overlap criterion by
\begin{equation}
\left(\frac{x}{\sqrt{\Lambda_1}+R_n}\right)^2+\left(\frac{y}{\sqrt{\Lambda_2}+R_n}\right)^2
+\left(\frac{z}{\sqrt{\Lambda_3}+R_n}\right)^2<1~,
\label{overlap}
\end{equation}
where $(x,y,z)$ here represent the coordinates of the vector joining the centers
of the sphere and ellipsoid.

In the event that a trial move results in a change in the number $\Delta N_{pn}$ 
of polymer-nanoparticle overlaps, then $\Delta U=\varepsilon\Delta N_{pn}$.
Thus, any displacement or rotation that reduces, or leaves unchanged, the number 
of overlaps is automatically accepted, while a move that creates new overlaps is 
accepted only with a probability equal to the Boltzmann factor for $\Delta U$.
For trial rotations, we define the orientation of a polymer by a unit vector 
${\bf u}$, aligned with the long ($\lambda_1$) axis of the ellipsoid
at polar angle $\theta$ and azimuthal angle $\phi$, 
and generate a new (trial) direction ${\bf u}'$ via
\begin{equation}
{\bf u}'=\frac{{\bf u}+\tau{\bf v}}{|{\bf u}+\tau{\bf v}|}~,
\label{rotation}
\end{equation}
where ${\bf v}$ is a unit vector with random orientation and $\tau$ is a
tolerance that determines the magnitude of the trial rotation~\cite{frenkel-smit2001}.
To confirm even sampling of orientations, we checked that histograms of 
$\cos\theta$ and $\phi$ for a free (i.e., uncrowded) polymer were flat.

A trial change in shape of an ellipsoidal polymer coil, from an old shape
$\lambda_{\rm old}$ to a new shape $\lambda_{\rm new}=\lambda_{\rm old}
+\Delta\lambda$, is accepted with probability 
\begin{equation}
{\cal P}_{\rm shape}(\lambda_{\rm old}\to\lambda_{\rm new}) = 
\min\left\{\frac{P_r(\lambda_{\rm new})}{P_r(\lambda_{\rm old})}
e^{-\beta\Delta U},~1\right\}~,
\label{shape-variation}
\end{equation}
where $\lambda\equiv(\lambda_1,\lambda_2,\lambda_3)$ collectively denotes
the eigenvalues and $P_r(\lambda)$ is the reservoir polymer shape distribution
[Eqs.~(\ref{Prlambda}) and (\ref{Eurich-Maass})].
Thus, a trial shape change is accepted with a probability equal to the Boltzmann factor
for the change in potential energy multiplied by the ratio of the new to the old
shape probabilities.  Through trial changes in gyration tensor eigenvalues, a polymer 
explores the landscape of possible shapes in the presence of crowders and evolves
toward a new equilibrium shape distribution. 

One MC step of a simulation consists of a trial displacement of every nanoparticle,
followed by a trial displacement, rotation, and shape change of every polymer. 
To maximize computational efficiency, we chose tolerances of $0.01~\sigma_n$ 
for trial displacements, $\tau=0.001$ for trial rotations, and for trial shape
(eigenvalue) changes, $\Delta\lambda_1=0.01$, $\Delta\lambda_2=0.003$,
and $\Delta\lambda_3=0.001$.
To facilitate extensions and portability of our simulation methods, we coded our 
MC algorithm in the Java programming language within the 
Open Source Physics library~\cite{osp-sip2006,osp2006}, exploiting the numerical and
visualization classes of the library.  The simulations thus run on any platform,
with a convenient graphical user interface, and so may have both scientific and
pedagogical value.

\subsection{Free-Volume Theory of Crowding}\label{theory} 
For the model polymer-nanoparticle mixtures described in Sec.~\ref{models},
Denton \etalia~\cite{denton-may2014} recently developed a free-volume theory,
which generalizes the theory of Lekkerkerker \etalia~\cite{lekkerkerker1992} 
from incompressible, spherical polymers to compressible, aspherical polymers.
To guide our choice of parameters, check for consistency, and test the theory,
we compare our simulation results with theoretical predictions.  As outlined
in Appendix~\ref{appendix}, the theory predicts a crowded-polymer shape 
probability distribution of the form
\begin{equation}
P(\lambda;\phi_n) = P_r(\lambda)\frac{\alpha(\lambda;\phi_n)}{\alpha_{\rm eff}(\phi_n)}~,
\label{Plambda}
\end{equation}
where the free-volume fraction $\alpha(\lambda;\phi_n)$ is the fraction of the
total volume accessible to a polymer, whose ellipsoidal shape is characterized by 
the eigenvalues $\lambda=(\lambda_1,\lambda_2,\lambda_3)$, amidst nanoparticles 
of volume fraction $\phi_n=n_n(4\pi/3)R_n^3$ (number density $n_n$), and
\begin{equation}
\alpha_{\rm eff}(\phi_n) \equiv \int{\rm d}\lambda\,P_r(\lambda)\alpha(\lambda;\phi_n)
\label{alphaeff}
\end{equation}
is an {\it effective} polymer free-volume fraction, expressed as an average of 
$\alpha(\lambda;\phi_n)$ over polymer shapes in the reservoir.
In practice, we adopt the ansatz for $P_r(\lambda)$ described in Sec.~\ref{penetrable-polymer}
and compute $\alpha(\lambda;\phi_n)$ by implementing the generalized 
scaled-particle theory of Oversteegen and Roth~\cite{oversteegen-roth2005}.
From Eq.~(\ref{Plambda}), the probability distribution for a single eigenvalue
is obtained by integrating over the other two eigenvalues.  For example,
\begin{equation}
P_1(\lambda_1;\phi_n) = \int_0^{\infty} d\lambda_2\,\int_0^{\infty} d\lambda_3\, P(\lambda;\phi_n)~.
\label{Plambda1}
\end{equation}
In calculating the rms radius of gyration [Eq.~(\ref{gyration-ratio})] 
and asphericity [Eq.~(\ref{asphericity})], mean values of functions of
eigenvalues $f(\lambda)$ are defined as averages with respect to $P(\lambda)$:
\begin{equation}
\la f\ra = \int d\lambda\, P(\lambda)f(\lambda)~.
\label{mean-lambda}
\end{equation}
In the next section, we present numerical results from MC simulations and 
free-volume theory that characterize the shapes of ideal polymers 
in crowded environments.

\begin{figure}
\begin{center}
\includegraphics[width=0.8\columnwidth]{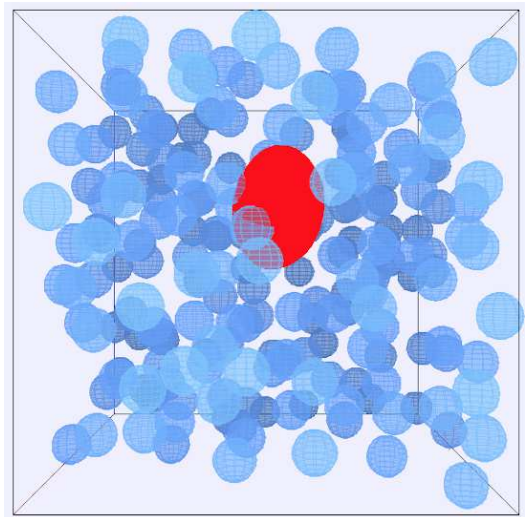}
\end{center}
\vspace*{-0.5cm}
\caption{
Snapshot of a simulation of $N_n=216$ nanoparticles (blue spheres) and one polymer
(red ellipsoid) in a cubic box.  The polymer rms radius of gyration in the reservoir
equals five times the nanoparticle radius ($q_r=5$).
}\label{fig-snapshot}
\end{figure}

\begin{figure}
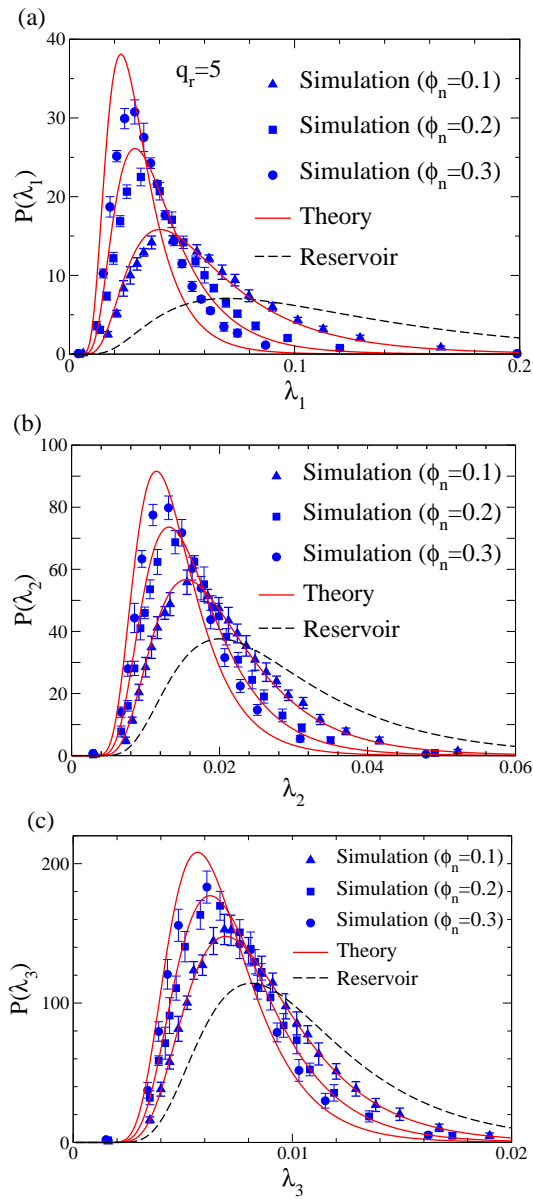

\includegraphics[width=0.8\columnwidth]{plambda1.q5.eps}
\includegraphics[width=0.8\columnwidth]{plambda2.q5.eps}
\includegraphics[width=0.8\columnwidth]{plambda3.q5.eps}
\vspace*{-0.2cm}
\caption{
Probability distributions for the eigenvalues (a) $\lambda_1$, (b) $\lambda_2$, 
and (c) $\lambda_3$ of the gyration tensor of a polymer coil, 
modeled as an ideal, freely-jointed chain.  Monte Carlo simulation data (symbols)
are compared with predictions of free-volume theory (solid curves) for a 
single ellipsoidal polymer, with rms radius of gyration in the reservoir equal to
five times the nanoparticle radius ($q_r=5$), amidst $N_n=216$ nanoparticles with 
volume fraction $\phi_n=0.1$ (triangles), 0.2 (squares), and 0.3 (circles).  
Also shown are the reservoir distributions (dashed curves), in the absence of 
nanoparticles ($\phi_n=0$).
}\label{fig-Plambda-q5}
\end{figure}

\begin{figure}
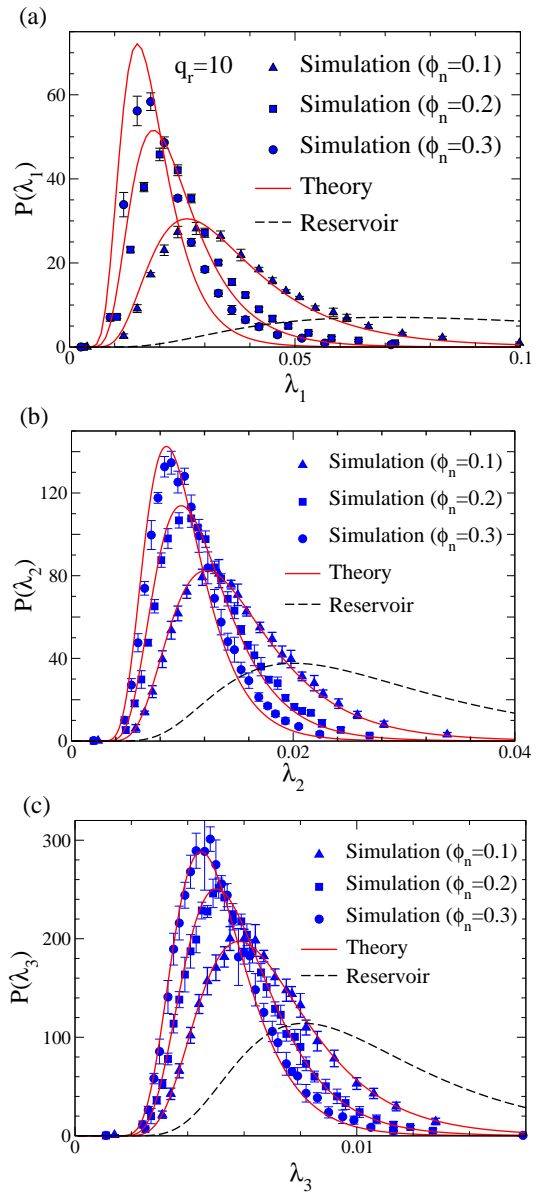

\includegraphics[width=0.8\columnwidth]{plambda1.q10.eps} 
\includegraphics[width=0.8\columnwidth]{plambda2.q10.eps}
\includegraphics[width=0.8\columnwidth]{plambda3.q10.eps}
\vspace*{-0.2cm}
\caption{
Same as Fig.~\ref{fig-Plambda-q5}, but for larger polymer ($q_r=10$).
Notice the changes in scale.
}\label{fig-Plambda-q10}
\end{figure}

\begin{figure}
\includegraphics[width=0.8\columnwidth]{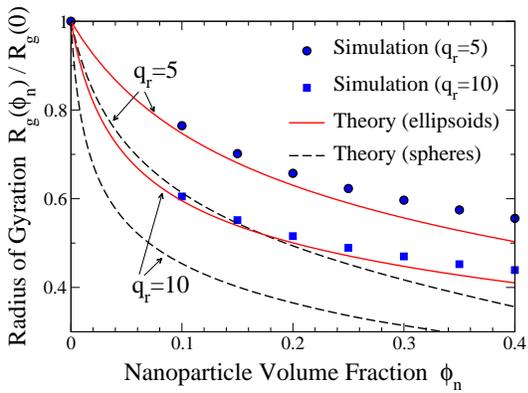}
\caption{
Root-mean-square radius of gyration of a polymer vs.~nanoparticle volume fraction
[Eq.~(\ref{gyration-ratio})]. 
Monte Carlo simulation data (symbols) are compared with predictions of 
free-volume theory for ellipsoidal polymer (solid curves) and spherical polymer
(dashed curves).  Results are shown for reservoir polymer-to-nanoparticle size ratio
$q_r=5$ (circles), $q_r=10$ (squares).  (Error bars are smaller than symbols.)
}\label{fig-rg}
\end{figure}

\begin{figure}
\includegraphics[width=0.8\columnwidth]{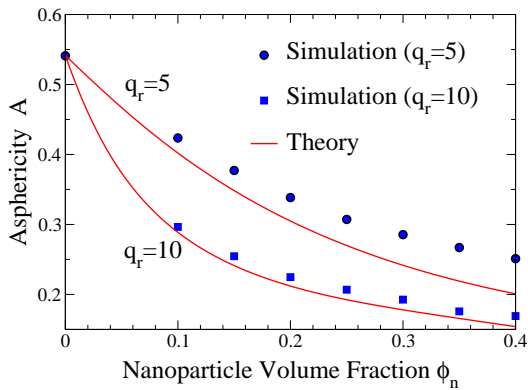}
\vspace*{-0.2cm}
\caption{
Asphericity of an ellipsoidal polymer vs.~nanoparticle volume fraction
[Eq.~(\ref{asphericity})]. 
Monte Carlo simulation data (symbols) are compared with predictions of 
free-volume theory (curves).  Results are shown for reservoir polymer-to-nanoparticle
size ratio $q_r=5$ (circles) and $q_r=10$ (squares).  (Error bars are smaller than symbols.)
As crowding increases, the polymer becomes more compact (less aspherical).
}\label{fig-asphericity}
\end{figure}

\begin{figure}
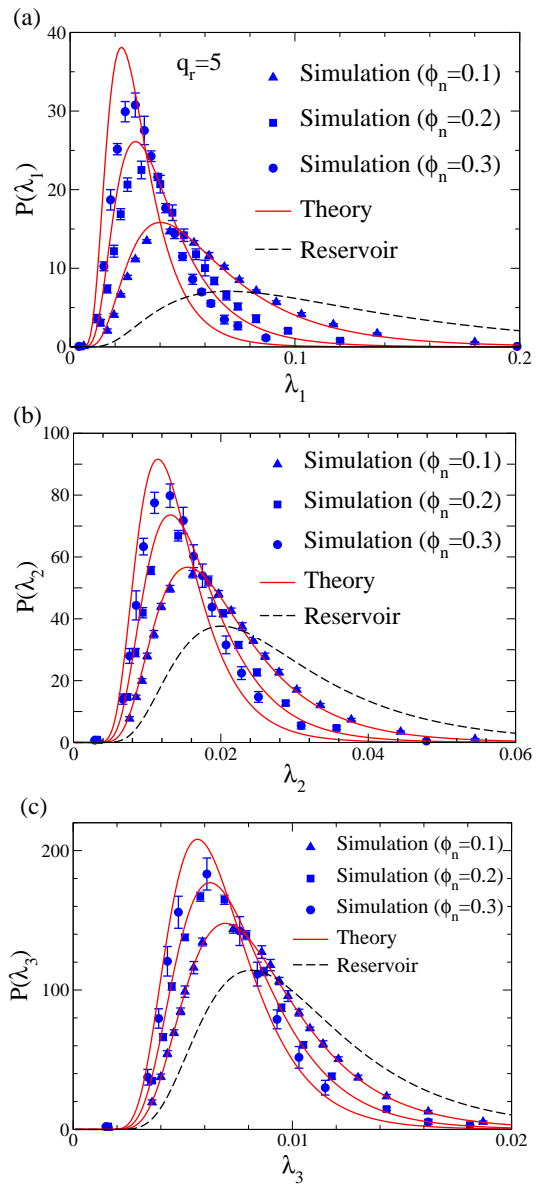

\begin{center}
\includegraphics[width=0.8\columnwidth]{plambda1.q5.np8.eps}
\includegraphics[width=0.8\columnwidth]{plambda2.q5.np8.eps}
\includegraphics[width=0.8\columnwidth]{plambda3.q5.np8.eps}
\vspace*{-0.2cm}
\caption{
Same as Fig.~\ref{fig-Plambda-q5}, but for higher polymer concentration 
($N_p=8$, $\phi_p\simeq 0.5$).
}\label{fig-Plambda-q5-phip05}
\end{center}
\end{figure}

\section{Results and Discussion}\label{results}
To investigate how the shapes of ideal polymers respond to crowding,
we simulated compressible, penetrable polymers, immersed in a fluid of smaller,
hard-sphere nanoparticles (protein limit), modeled as described 
in Sec.~\ref{models}, and using the MC method outlined in Sec.~\ref{simulation}.
Confining the system to a cubic box of fixed size with periodic boundary conditions
applied to opposite faces, we initialized the nanoparticles on the sites of a cubic
lattice and the polymers at interstitial sites.  For illustration, a snapshot
of the simulation cell is shown in Fig.~\ref{fig-snapshot}.  

Each run consisted of an initial equilibration stage of $5\times 10^4$ MC steps, 
followed by a data collection stage of $10^7$ steps.  We monitored the total overlap
energy and shape distributions and confirmed that the averages of these diagnostics 
were stable after the equilibration stage.  Our results represent averages over $10^4$
independent configurations (spaced by intervals of $10^3$ steps) from each of five
independent runs (total of $5\times 10^4$ configurations), with statistical 
uncertainties computed from standard deviations of the five runs.
Most of our simulations were performed for systems of $N_n=216$ nanoparticles.
To rule out finite-size effects, however, we repeated several runs for larger systems
(up to $N_n=1728$) and confirmed that the results are independent of system size
to within statistical fluctuations.

Figure~\ref{fig-Plambda-q5} shows the probability distributions for the eigenvalues
of the gyration tensor, representing the shape of the best-fit ellipsoid,
for one polymer amidst $N_n=216$ nanoparticles, with the reservoir 
rms radius of gyration equal to five times the nanoparticle radius ($q_r=5$).
At this large size ratio, our approximation for the polymer-nanoparticle overlap 
criterion [Eq.~(\ref{overlap})] is quite accurate.
With increasing nanoparticle volume fraction, from $\phi_n=0$ (reservoir) to 
$\phi_n=0.3$, the shape distributions progressively shift toward smaller eigenvalues,
reflecting compression of the polymer along all three principle axes.  
The greatest fractional shift occurs, however, in the two largest eigenvalues 
($\lambda_1$ and $\lambda_2$), implying that the best-fit ellipsoids tend to become
less elongated.

Figure~\ref{fig-Plambda-q10} shows the probability distributions for a polymer 
twice as large ($q_r=10$).  Doubling the size ratio, while still avoiding
significant finite-size effects, required doubling the simulation box length,
and thus increasing eight-fold the number of nanoparticles ($N_n=1728$).
As a rough guide, the simulation box must be large enough that the long axis 
of the polymer cannot span a significant fraction of the box length.  Otherwise,
correlations between a polymer and its own images can cause spurious effects.
To minimize computational time for the larger system, we reduced the run length 
to $10^6$ MC steps, without a significant change in results.
Our runs of $10^7$ steps proved, therefore, to be conservatively long.

For the same nanoparticle concentration, the shape distributions of the larger polymer
are considerably more shifted relative to the reservoir distributions.  This trend is 
easily explained by considering the average free energy cost $\bar E_{pn}$ of 
polymer-nanoparticle overlaps.  Neglecting correlations, the average number of 
overlaps scales as $\phi_n q^3$, while the penetration energy scales as $q^{-1}$.  
Thus, the average overlap energy scales as $\bar E_{pn}\sim\phi_nq^2$, i.e., 
the crowding effect increases with the square of the size ratio. 

Also shown in Figs.~\ref{fig-Plambda-q5} and \ref{fig-Plambda-q10} are the shape
distributions predicted by the free-volume theory, described in Sec.~\ref{theory} 
and the Appendix.  In this limit of dilute polymer concentration, theory and simulation
are evidently in close agreement at lower nanoparticle concentrations.  As the polymer
becomes increasingly crowded, however, slight quantitative deviations emerge,
particularly for the largest eigenvalue $\lambda_1$ of the gyration tensor at $q_r=5$.  
These small deviations result from the mean-field theory's neglect of polymer-nanoparticle
correlations and from approximations inherent in scaled-particle theory. 

From the polymer shape (eigenvalue) distributions, we have computed the rms radius 
of gyration [Eq.~(\ref{gyration-ratio})] and asphericity [Eq.~(\ref{asphericity})]
of a single crowded polymer as functions of nanoparticle concentration.
As shown in Figs.~\ref{fig-rg} and \ref{fig-asphericity}, an ideal polymer responds
to crowding by contracting in size (decreasing $R_g$) {\it and} becoming more 
spherical in shape (decreasing $A$).  Thus, with increasing nanoparticle volume fraction,
the polymer progressively compactifies.
Increasing the size ratio from $q_r=5$ to $q_r=10$ enhances the crowding effect,
for reasons explained above, the polymer becoming even smaller and more spherical 
for a given nanoparticle concentration.  

Figures~\ref{fig-rg} and \ref{fig-asphericity} also show that the free-volume theory
again accurately captures the trends in size and shape.  Nevertheless, small gaps 
between theory and simulation are apparent, and these quantitative deviations grow
with increasing nanoparticle concentration.  The theory's slight, but consistent,
underprediction of both $R_g$ and $A$ is due mainly to the underprediction of $\lambda_1$.
To emphasize the distinction between ellipsoidal and spherical polymer models,
Fig.~\ref{fig-rg} also shows, for comparison, free-volume theory predictions for 
a spherical, compressible polymer model~\cite{denton-schmidt2002,lu-denton2011}.
Clearly ellipsoidal polymers, being free to distort their shape, have significantly
larger radii of gyration in crowded environments than polymers that are constrained
to remain spherical.

To explore crowding at higher polymer concentrations, we increased the polymer 
volume fraction to $\phi_p\equiv n_p(4\pi/3)(R_g^r)^3\simeq 0.5$, with $N_p=8$ polymers
now sharing the simulation box with $N_n=216$ nanoparticles at size ratio $q_r=5$.  
These conditions actually place the system in a part of the phase diagram 
that is thermodynamically unstable toward polymer-nanoparticle 
demixing~\cite{moncho-jorda2003,vanduijneveldt2006}.  Bulk phase separation is prevented
only by the constraints of the NVT ensemble and the relatively small system size.
As illustrated in Fig.~\ref{fig-Plambda-q5-phip05}, the simulated shape distributions
do not substantially differ from those for a single polymer (Fig.~\ref{fig-Plambda-q5}).
Interestingly, this behavior differs from that observed in simulations of the spherical,
compressible, ideal polymer model~\cite{lu-denton2011} in the colloid limit ($q_r=1$),
where polymer compression reversed with increasing crowding.  This reversal, caused by
polymer clustering and shielding -- a correlation effect neglected by the mean-field
free-volume theory -- is not observed here in the protein limit.  

In closing this section, we briefly discuss the relation of our approach to 
experiments and other modeling approaches.
Recent studies that applied small-angle neutron scattering to polystyrene chains 
in the presence of various molecular crowding agents~\cite{kramer2005a,kramer2005b,
kramer2005c}, and to deuterated PEG amidst the polysaccharide crowder
Ficoll 70~\cite{longeville2009,longeville2010}, reported substantial crowding-induced
polymer compression.  Although the polymers in these experiments were nonideal
and relatively close in size to the crowders, our results for ideal polymers and
larger size ratios are at least qualitatively consistent with these observations.

The role of crowding in native-denatured transitions of real polypeptides 
was recently modeled by Minton~\cite{minton2005}.  Applying an effective two-state
model of proteins~\cite{minton2000}, Minton calculated excluded-volume interactions
between unfolded proteins and macromolecular cosolutes, modeled as hard spheres or rods.
Taking as input the radius of gyration probability distributions of four real proteins,
computed by Goldenberg~\cite{goldenberg2003} via Monte Carlo simulations that include
steric interactions between nonadjacent amino acid residues, Minton calculated
chemical potentials and radii of gyration of unfolded proteins as a function
of cosolute concentration.  He concluded that long-range intramolecular steric 
interactions significantly increase the radii of gyration of unfolded polypeptides
in crowded environments.  Our approach can potentially complement Minton's by 
incorporating knowledge of both the size {\it and} shape of the uncrowded polymer.

\section{Conclusions}\label{conclusions}
In summary, we have investigated the influence of crowding on polymer shapes in a
coarse-grained model of polymer-nanoparticle mixtures.  The ideal polymer coils 
are modeled here as effective ellipsoids that fluctuate in shape according to the 
probability distributions of the eigenvalues of the gyration tensor of a random walk.
The nanoparticles are modeled as hard spheres that can penetrate the polymers with
a free energy penalty varying inversely with the polymer-to-nanoparticle size ratio $q_r$.
For this model, we performed both Monte Carlo simulations, incorporating novel
trial moves that change the polymer shape, and free-volume theory calculations. 
In the protein limit, for size ratios of $q_r=5$ and 10, we computed the 
shape distributions, radius of gyration, and asphericity of ideal polymers
induced by crowding of hard-sphere nanoparticles.  Relative to uncrowded polymers, 
we observed significant shifts in polymer shape, which grow with increasing 
nanoparticle concentration and size ratio.  Our results demonstrate that ideal
polymers become more compact when crowded by smaller, hard nanoparticles,
in good agreement with predictions of free-volume theory.
The methods and results presented here significantly extend the scope of 
previous studies of colloid-polymer mixtures in which the polymers were 
modeled as compressible spheres~\cite{denton-schmidt2002,lu-denton2011}.

For future work, we envision several intriguing directions in which our approach
may be extended.  While the present paper focuses on the influence of nanoparticles
on polymers, one could, conversely, study the impact of polymers on effective
interactions between nanoparticles.  In particular, by simulating a pair of 
nanoparticles in a bath of shape-fluctuating polymers, the depletion-induced 
potential of mean force between nanoparticles could be computed and compared
with simulations of more microscopic models~\cite{bolhuis2003}, as well as with 
predictions of polymer field theory~\cite{eisenriegler2003} and density-functional
theory~\cite{forsman2009,forsman2014}, in the protein limit.

Our model can be refined by replacing the step-function polymer-nanoparticle
overlap energy profile with a more realistic, continuous profile based on
the monomer density profile~\cite{eurich-maass2001} or on
molecular simulations~\cite{pelissetto-hansen2006}.  Furthermore, by replacing
the shape distribution of an ideal (non-self-avoiding) random walk with that of
a nonideal (self-avoiding) walk~\cite{lhuilier1988,sciutto1996,schaefer1999}, 
the model can be extended from ideal polymers in theta solvents to real polymers
in good solvents.  Such extensions can include biopolymers in aqueous solutions,
such as unfolded proteins, whose persistence lengths can be sensitive to 
excluded-volume interactions~\cite{minton2005}, and whose uncrowded 
size distributions can be independently computed~\cite{goldenberg2003}.
For a single biopolymer in a crowded environment, our computational methods 
can be directly applied, given as input the requisite shape 
distribution~\cite{denesyuk2011,kudlay-thirumalai2012,denesyuk2013}.
Simulating solutions of multiple self-avoiding polymers would require 
incorporating polymer-polymer interactions~\cite{frenkel-mulder1985,allen1993}.
It is important to note, however, that our Monte Carlo approach, while efficiently
sampling polymer conformations, does not accurately represent time scales 
for distinct molecular motions -- diffusion, rotation, and shape fluctuations.
Therefore, our methods, while finding equilibrium shapes of crowded polymers,
cannot describe dynamical processes, such as folding and unfolding. 

Beyond adding realism to the polymer model, our approach can also be extended
to mixtures of polymers with nonspherical~\cite{kudlay-thirumalai2012} or 
charged~\cite{denton-schmidt2005,tuinier2005} crowders, or to other crowded environments,
such as confinement within a vesicle~\cite{may-kroll2013}, or two-dimensional
confinement, e.g., of DNA adsorbed onto lipid membranes~\cite{fang1997,maier2000}.
Finally, for all of these systems, it would be interesting to explore the influence 
of polymer shape degrees of freedom on bulk thermodynamic properties, including the
demixing transition between polymer-rich and polymer-poor phases, by implementing
our Monte Carlo methods in either the Gibbs ensemble~\cite{lu-denton2011} or the 
grand canonical ensemble~\cite{vink-horbach-jcp2004,vink-horbach-jpcm2004}.

\acknowledgments
We thank Sylvio May, Emmanuel Mbamala, Ben Lu, Matthias Schmidt, and James Polson
for discussions.  This work was supported by the National Science Foundation 
(Grant No.~DMR-1106331) and by the Donors of the American Chemical Society
Petroleum Research Fund (Grant No.~PRF 44365-AC7).

\appendix
\section{Free-Volume Theory}\label{appendix}
Here, we outline in greater detail the theory sketched in Sec.~\ref{theory}.
In the semi-grand ensemble, a fixed number $N_n$ of nanoparticles are confined 
to a volume $V$, while the polymers can exchange with a reservoir of polymer 
that maintains constant polymer chemical potential $\mu_p$ in the system.  
At a given temperature, the thermodynamic state is characterized by the nanoparticle 
number density, $n_n=N_n/V$, and the polymer number density in the reservoir,
$n_p^r\propto\exp(\beta\mu_p)$ (ideal polymer).
The polymer number density in the system, $n_p=N_p/V$, which depends on the 
nanoparticle density, is determined by chemical equilibrium between the 
system and reservoir.

The free-volume theory, a generalization of the theory first proposed by
Lekkerkerker \etalia~\cite{lekkerkerker1992} for the AOV model of colloid-polymer mixtures,
can be derived by separating the Helmholtz free energy density,
$f=f_{\rm id}+f_{\rm ex}$, into an ideal-gas contribution $f_{\rm id}$
and an excess contribution $f_{\rm ex}$ due to interparticle interactions.
The excess free energy density consists of a hard-sphere nanoparticle contribution
$f_{\rm hs}(\phi_n)$
and a polymer contribution $f_p$, which depends on polymer-nanoparticle interactions.  
In a mean-field approximation, the polymer excess free energy density
is equated to that of ideal polymers confined to the free volume
(not excluded by the nanoparticles).

For shape-fluctuating polymers, the free energy must be averaged over 
shape degrees of freedom and supplemented by a conformational free energy.
Assuming that a polymer of a given shape (i.e., eigenvalues $\lambda$) 
has the same conformational entropy in the system as in the reservoir,
namely $k_B\ln P_r(\lambda)$, the polymer excess free energy density is 
approximated by
\begin{equation}
\beta f_p(\phi_n,\phi_p) = -n_p\int{\rm d}\lambda\,P(\lambda;\phi_n)
\ln[P_r(\lambda)\alpha(\lambda;\phi_n)]~,
\label{fexp2}
\end{equation}
where $P(\lambda;\phi_n)$ and $\alpha(\lambda;\phi_n)$ are the probability distribution
and free-volume fraction, respectively, of polymer coils of shape $\lambda$ 
amidst nanoparticles of volume fraction $\phi_n\equiv (4\pi/3)n_nR_n^3$.
The ideal-gas free energy density is given exactly by
\begin{eqnarray}
\beta f_{\rm id}(\phi_n,\phi_p)&=&
n_p\int{\rm d}\lambda\,P(\lambda;\phi_n)\{\ln[\phi_p P(\lambda;\phi_n)]-1\}
\nonumber\\[1ex]
&+&n_n\left(\ln\phi_n-1\right)~, 
\label{fid}
\end{eqnarray}
where $\phi_p\equiv (4\pi/3)n_p(R_g^r)^3$ is the effective polymer volume fraction
in the system and $R_g^r$ is the rms radius of gyration in the reservoir.

Equating chemical potentials of ideal polymers of a given shape 
in the system and reservoir now implies 
\begin{equation}
n_p(\phi_n)P(\lambda;\phi_n)=n_p^rP_r(\lambda)\alpha(\lambda;\phi_n)~.
\label{nprp}
\end{equation}
Integrating over $\lambda$ and using the normalization of 
$P(\lambda;\phi_n)$ yields
\begin{equation}
n_p(\phi_n)=n_p^r\alpha_{\rm eff}(\phi_n)~,
\label{np}
\end{equation}
where $\alpha_{\rm eff}$ is an effective polymer free-volume fraction, 
\begin{equation}
\alpha_{\rm eff}(\phi_n) \equiv \int{\rm d}\lambda\,P_r(\lambda)\alpha(\lambda;\phi_n)~,
\label{alphaeff-appendix}
\end{equation}
defined as an average of the free-volume fraction over polymer shapes in the reservoir.
The corresponding shape distribution of crowded polymers is 
\begin{equation}
P(\lambda;\phi_n) = P_r(\lambda)\frac{\alpha(\lambda;\phi_n)}{\alpha_{\rm eff}(\phi_n)}~.
\label{Plambda-appendix}
\end{equation}
Note that in the dilute nanoparticle limit ($\phi_n\to 0$), the free-volume fraction
$\alpha\to 1$ and the shape distribution reduces to that of the reservoir:
$P(\lambda)\to P_r(\lambda)$.
Collecting the various contributions, the total free energy density
may be expressed as
\begin{eqnarray}
\beta f(\phi_n,\phi_p^r) &=& n_n\left(\ln\phi_n-1\right) + \beta f_{\rm hs}(\phi_n) 
\nonumber\\[1ex]
&+& n_p^r\alpha_{\rm eff}(\phi_n)(\ln\phi_p^r-1)~,
\label{ftot}
\end{eqnarray}
where now $\phi_p^r\equiv (4\pi/3)n_p^r(R_g^r)^3$ is the effective polymer 
volume fraction in the reservoir.

For the polymer free-volume fraction, we adopt the accurate geometry-based
approximation of Oversteegen and Roth~\cite{oversteegen-roth2005},
which generalizes scaled-particle theory~\cite{lebowitz1964} from spheres
to arbitrary shapes by using fundamental-measures density-functional
theory~\cite{rosenfeld1989,rosenfeld1997,schmidt2000}
to separate thermodynamic properties of the crowders (nanoparticles) from
geometric properties of the depletants (polymers).  
The result is
\begin{equation}
\alpha(\lambda;\phi_n) = (1-\phi_n)\exp[-\beta(pv_p+\gamma a_p+\kappa c_p)]~,
\label{alpha-fmt}
\end{equation}
where $p$, $\gamma$, and $\kappa$ are the bulk pressure, surface tension at a 
planar hard wall, and bending rigidity of the nanoparticles, while $v_p$, $a_p$, 
and $c_p$ are the volume, surface area, and integrated mean curvature of a polymer.
For a spherical polymer, $v_p=(4\pi/3)R_p^3$, $a_p=4\pi R_p^2$, and $c_p=R_p$.
A general ellipsoid polymer, with principal radii $R_1$, $R_2$, $R_3$, has volume 
$v_p=(4\pi/3)R_1R_2R_3$, while $a_p$ and $c_p$ are numerically evaluated
from the principal radii.
The thermodynamic properties of hard-sphere nanoparticles are accurately
approximated by the Carnahan-Starling 
expressions~\cite{oversteegen-roth2005,hansen-mcdonald2006}:

\begin{eqnarray}
\beta f_{{\rm hs}}&=&n_n\frac{\phi_n(4-3\phi_n)}{(1-\phi_n)^2}
\nonumber \\[0.5ex]
\beta p&=&\frac{3\phi_n}{4\pi R_n^3}\frac{1+\phi_n+\phi_n^2-\phi_n^3}{(1-\phi_n)^3}
\nonumber \\[0.5ex]
\beta\gamma&=&\frac{3}{4\pi R_n^2}\left[\frac{\phi_n(2-\phi_n)}{(1-\phi_n)^2}
+\ln(1-\phi_n)\right] \nonumber \\[0.5ex]
\beta\kappa&=&\frac{3\phi_n}{R_n(1-\phi_n)}~.
\label{CSthermo}
\end{eqnarray}


\bibliography{cpmresub}

\end{document}